\documentclass[11pt]{article}
\usepackage[paperwidth=8.5in, paperheight=11.in,
body={6.5in,9.5in},top=1.25in,bottom=1.in,headsep=0.22in,headheight=0.25in,
left=1.in,right=1.in]{geometry} 
\usepackage{amssymb}
\usepackage{amsmath}
\usepackage{wrapfig}
\usepackage{subcaption}
\usepackage{array}
\usepackage[pdftex]{graphicx}
\usepackage{fancyhdr}
\usepackage{xspace}
\usepackage[numbers,compress]{natbib}
\usepackage[compact,small]{titlesec}
\usepackage{enumitem}
\usepackage{mdwlist}
\usepackage[hidelinks]{hyperref}
\usepackage{sidecap} 
\usepackage[svgnames]{xcolor}
\usepackage{mdframed}
\usepackage{tcolorbox}
\usepackage{longtable}
\usepackage{caption}
\usepackage{pdflscape}
\usepackage{multirow}
\usepackage{hhline}
\usepackage[table]{colortbl}% http://ctan.org/pkg/xcolor

\tcbuselibrary{breakable}

\usepackage{sectsty}
\usepackage[final]{pdfpages}

\newcommand{\squishlist}{
 \begin{list}{$\bullet$}
  { \setlength{\itemsep}{-2pt}
     \setlength{\parsep}{3pt}
     \setlength{\topsep}{3pt}
     \setlength{\partopsep}{0pt}
     \setlength{\leftmargin}{1.5em}
     \setlength{\rightmargin}{0em}
     \setlength{\labelwidth}{1em}
     \setlength{\labelsep}{0.5em} } }
\newcommand{\squishend}{
  \end{list}  }

\definecolor{MyGray}{rgb}{0.9,0.9,0.9}


\begin{document}

%\lhead[]{ NNH24ZDA001N-LUCYL4PSP} %
%\rhead[]{The Dust Environment of Trojans} %
%\cfoot[{\thepage}]{{\thepage}}

%% active preamble
%\pagenumbering{roman} %
\newcommand{\Section}[1]{\clearpage\section{#1} \setcounter{page}{1}}
\newcommand{\captionbaseline}{\renewcommand{\baselinestretch}{0.9}} % SPECIFY BASELINES HERE
\newcommand{\mainbaseline}{\renewcommand{\baselinestretch}{0.935}} % SPECIFY BASELINES HERE

%%%%%%%%%%%%%%%%%%%%%%%%%%%%%%%%%%%%%%%%%%%%%%%%%%%%%%%%%%%%%%%%%%%%%%%%%%%%%%
%% and the document actually starts here

\raggedright
\LARGE
SPIRE HeRS/HeLMS Combined SHIM Maps \linebreak
\normalsize

\textbf{Authors:}
Michael Zemcov$^{1,2}$, Richard Feder$^{3}$, Ryan Wills$^{1}$
 \linebreak						
 \small $^{1}$\textit{mbzsps@rit.edu, Rochester Institute of
 Technology, 1 Lomb Memorial Dr., Rochester, NY 14623, USA}; $^{2}$\textit{Jet
 Propulsion Laboratory, 4800 Oak Grove Drive, Pasadena, CA 91109,
 USA}; $^{3}$\textit{Berkeley Center for Cosmological Physics, University of California, Berkeley, CA 94720, USA}
\normalsize
 
\vspace{20pt}
\justifying

We have regenerated \textit{Herschel}-SPIRE maps covering 360 square degrees
near the celestial equator.  These are the largest extragalactic surveys
designed to 
overlap with cosmic microwave background legacy fields mapped at sub-mm
wavelengths.  We provide documentation detailing their construction and
use.  The maps are available on zenodo as \href{https://doi.org/10.5281/zenodo.13352296}{10.5281/zenodo.13352296}.

\section*{Data Description}

The repository contains sub-mm wavelength sky images of the combined
\textit{Herschel} Redshift Survey (HeRS, sometimes also the
\textit{Herschel} Stripe 82 Survey; \cite{Viero2014}) and \textit{Herschel} Large
Mode Survey (HeLMS; \cite{Oliver2012}) fields as observed by the SPIRE
photometer \cite{Griffin2010} on the \textit{Herschel Space
  Observatory} \cite{Pilbratt2010} at 250, 350, and 500 $\mu$m.  These
regions were selected to overlap with the SDSS-Stripe82
\cite{Abazajian2009} and ACT-CMB \cite{Das2011} fields in a contiguous
fashion.  The maps cover areas of approximately 360 square degrees at
6, 8.3, and 12 arcsec pixel$^{-1}$.

The data in this repository improve on previous analyses of these
data, which were heterogeneously analyzed, served on non-permanent
archives, and occasionally had errors discovered by the user
community.  The previous analyses include:
\squishlist
\item HeRS survey maps constructed ~2012 with the SHIM map maker, described in \cite{Viero2014};
\item HeLMS maps constructed ~2015 with the SANEPIC map maker, described in \cite{Asboth2016};
\item HeRS+HeLMS+XMM maps constructed ~2016 with the SHIM map maker,
  described in \cite{Help2016}; and
\item HeRS and HeLMS individual maps constructed ~2019 with the CADE
  map maker, described in \cite{Paradis2012}.
  \squishend 
While all of these maps are currently circulating in the community, we expect their availability to wane as time goes on.  Further, in many cases the detailed reduction steps and provenance of the served products are unclear.  In this repository we have uploaded vetted and error-free fits files constructed in a consistent way as follows.

To start, we use the {\sc Herschel Interactive Processing Environment} (HIPE, version 15.0.1) and the most recent calibration files (version 14.3) to reduce the level zero observations to level one data products.  The code and calibration tree are the last released by the \textit{Herschel} team, and we do not expect improvements or updates to them in the future.  For the products available in this download, our processing consists of the standard scanline calibration pipeline for SPIRE that comes pre-packaged in HIPE (Photometer Large Map Pipeline) with a few changes. The pipeline converts the detector timelines from products in engineering units (satellite pointing and voltages) to science products (sky pointing and flux units) with corrections for cross-talk, signal jumps, glitches, cooler burps, the low pass filter response, and the bolometer time response. We chose to turn off the temperature drift correction as we found that the SHIM map-maker performs better without it and we opt to use the sigma-kappa deglitcher rather than the wavelet deglitcher.  We also do not perform any destriping or baseline subtraction of the scan lines with HIPE.  The HIPE script is available as part of this download for inspection.

We apply this HIPE reduction to the 33 observation IDs listed in the
zenodo record description (21 from proposal ID {\sc OT2\_mviero\_2}
\cite[HerS][]{Viero2013} and 12 from proposal ID {\sc GT2\_mviero\_1} \cite[HeLMS][]{Viero2011}), which are retrieved with HIPE from the \textit{Herschel} Science Archive (\href{http://archives.esac.esa.int/hsa/whsa/}{HSA}).

%\begin{table}
%Observation IDs
%1342247220
%1342247993
%1342247994
%1342247995
%1342247996
%1342247997
%1342247998
%1342248000
%1342248001
%1342248491
%1342248492
%1342248493
%1342248494
%1342248495
%1342248496
%1342248497
%1342248498
%1342248499
%1342248500
%1342249103
%1342249105
%1342234749
%1342236232
%1342236234
%1342236240
%1342237550
%1342237553
%1342237563
%1342238251
%1342246580
%1342246632
%1342247216
%1342257362

The level one observations are saved as HIPE-format fits files that
are input into the map-maker as follows.  Spatial maps of the
astrophysical emission are produced using the SHIM map maker
(\cite{Levenson2010}, with additional processing details provided in
\cite{Viero2013a}), which has been found to out-perform other map
makers in terms of retaining fidelity on all angular scales
\cite{Xu2014}.  SHIM is a \textit{Herschel}-SPIRE specific implementation of an
iterative baseline removal and detector noise weighting algorithm
originally described in \cite{Fixsen2000}.  The data are processed
scan-wise, i.e.~baselines and weights are calculated per one pass of
the photometer across a region of sky.  A noise map is created by
propagating detector noise as estimated by the variance of the
residuals and computing the weighted inverse sum in the map pixels.
This detector noise propagation methodology gives a better measure of
noise variations across the map and is more robust when the number of
samples per pixel is low and the standard deviation becomes a poor
estimate of the statistical error in the measurement.  Also included
in the fits files are exposure and flag information.  The images are
in units of [Jy/beam], which can be converted to units of surface
brightness in [MJy/Sr] by applying the multiplicative conversion
factors 90.646, 51.181, and 23.580 at 250, 350, and 500 microns,
respectively.  The error maps are reported as the square root of the
variance of the statistical noise in each pixel, and do not include
the (large) effect of confusion noise in these maps \cite[see
e.g.][]{Nguyen2010}.  Though we do not explicitly provide map transfer
function estimates, the transfer function is 1\% to angular scales of
$\sim 1$ degree, and thereafter falls exponentially with a 3dB point
of $\sim 6$ degrees.

The FITS files provided in this repository are labeled with the SPIRE
band names PSW, PMW, and PLW, which correspond to the 250, 350, and
500 $\mu$m channels, respectively. Each FITS file contains a header
data unit list (HDU, HDUL) with a primary HDU containing only very
basic header info and four image HDUs with more extensive header info
and the image data. Table 1 gives basic information on the HDUL. 

\begin{table}
  \caption{FITS extension descriptions.}
  \centering
  \begin{tabular}{|l|c|c|c|c|}
    \hline
    Ext number &	Name	& Type	& Format	& Units  \\ \hline
0	& PRIMARY	& PrimaryHDU	& N/A	& N/A \\ 
1	& image	& ImageHDU	& float64	& Jy/Beam \\ 
2	& error	& ImageHDU	& float64	& Jy/beam \\ 
3	& exposure	& ImageHDU	& float64	& seconds \\
    4	& mask	& ImageHDU	& int32	& N/A \\ \hline
  \end{tabular}
\end{table}

The ``image'' extension contains the science mosaics calibrated in $F_{\nu}$, the ``error'' extension contains the errors corresponding to the science mosaic (and are derived as described above), the ``exposure'' extension is the effective number of seconds of integration in each map pixel, and finally the ``mask'' extension contains the image mask, which is currently an array of zeros as no masking has been applied to these images. Each image extension also contains information to reference to the world coordinate system.

%\printbibliography

% References
%\bibliographystyle{unsrtnat}
\bibliography{zenodo_desc4arxiv}

%\vfill\eject

\end{document}